\documentclass[aps,twocolumn,showpacs,floats]{revtex4}
\usepackage{graphicx}
\usepackage{dcolumn}
\usepackage{bm}
\usepackage{color}
\usepackage{amssymb}

\begin{document}

\author{S. Pilati}
\affiliation{Theoretische Physik, ETH Zurich, 8093 Zurich, Switzerland}
\affiliation{The Abdus Salam International Centre for Theoretical Physics, 34014 Trieste, Italy}
\author{M. Troyer}
\affiliation{Theoretische Physik, ETH Zurich, 8093 Zurich, Switzerland}

\title{Bosonic Superfluid-Insulator Transition in Continuous Space }

\begin{abstract} 
We investigate the zero-temperature phase diagram of interacting Bose gases in the presence of a simple cubic optical lattice, going beyond the regime where the mapping to the single-band Bose-Hubbard model is reliable.
Our computational approach is a new hybrid quantum Monte Carlo method which combines algorithms used to simulate homogeneous quantum fluids in continuous space with those used for discrete lattice models of strongly correlated systems.
We determine the critical interaction strength and optical lattice intensity where the superfluid-to-insulator transition takes place, considering also the regime of shallow optical lattices and strong interatomic interactions.
The implications of our findings for the  supersolid state of matter are discussed.
\end{abstract}

\pacs{67.85.Hj, 03.75.Hh}
\maketitle 

In quantum many-body systems, the interplay between interparticle interactions and commensurability effects in external periodic potentials can induce a transition from a conducting (or superconducting) to an insulating phase, where particles are localized around the minima of the periodic potential. This phenomenon (known as the Mott transition) is responsible for the insulating behavior of several transition metal compounds~\cite{mott1990metal}. 
Recently, the Mott transition has been induced in a controlled way in both bosonic~\cite{greiner2002quantum} and fermionic~\cite{joerdens2008mott,schneider2008metallic} atomic gases trapped in optical lattices, which are being used as quantum emulators to explore the intriguing properties of strongly correlated materials~\cite{lewenstein2007ultracold}.

Previous theoretical studies of cold atomic gases in optical lattices are based on single-band lattice models, such as the Bose- or Fermi-Hubbard models~\cite{jaksch1998cold}. These theories are limited to the regime of large optical lattice intensity - since excited Bloch bands are neglected~\cite{jaksch1998cold} -  and weak interactions, since the true interatomic potential is replaced by a nonregularized $\delta$~function, which corresponds to the first Born approximation for the scattering problem.
The role of higher Bloch bands has been considered in Refs. \onlinecite{luehmann,bissbort} (by using the nonregularized pseudopotential), resulting in effective single-band models with density-dependent Hubbard parameters or many-body interactions.
The description of interactions beyond the first Born approximation is a challenge, since more accurate pseudopotentials such as the regularized $\delta$~function induce virtual transitions to an extremely large number of Bloch bands~\cite{buechler2010microscopic}.
The validity of the single-band approximation has been questioned on an even more profound level by Anderson~\cite{anderson2009gross,anderson2011ground}, in particular, concerning the existence of a bosonic Mott insulator. One central issue is the presence of nodes in the ground-state wave function of the bosonic Mott phase in the single-band Hubbard model, since this is built on the basis of orthogonal -and, hence, nonpositive definite - Wannier functions. Anderson further argues that when all Bloch bands are included the superfluid component is always finite, implying that the solid phase is a supersolid.

\begin{figure}[tb]
\begin{center}
\includegraphics[width=\columnwidth]{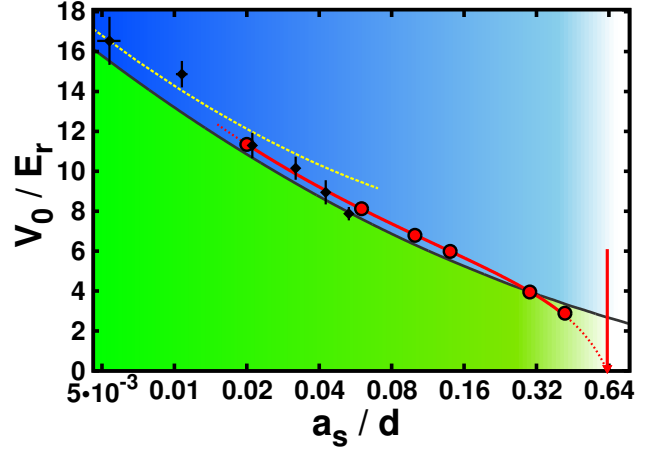}
\caption{(color online). Zero-temperature phase diagram as a function of the optical lattice intensity $V_0/E_\mathrm{r}$ and the $s$-wave scattering length $a_\mathrm{s}/d$ at unit filling. $E_\mathrm{r}$ is the recoil energy and $d$ is the optical lattice periodicity. The red points indicate the transition from the superfluid (green region) to the insulating phase (blue region) for hard-sphere particles. The solid gray curve is the phase boundary of the single-band Bose-Hubbard model~\cite{capogrosso2007phase} (for a nonregularized $\delta$-function interaction~\cite{jaksch1998cold}). The dashed yellow curve is the multiorbital model of Ref.~\cite{luehmann} (obtained within the mean-field approximation). The black diamonds are the experimental results of Ref.~\cite{mark}. The  solid red curve is a guide to the eye, and the dashed red curve is an extrapolation towards the freezing point without an optical lattice (red vertical arrow). See the text for other scenarios that are possible in this regime.}
\label{fig1}
\end{center}
\end{figure}

In this Letter, we map out the zero-temperature and continuous-space phase diagram of a Bose gas with short-range interactions in a simple cubic optical lattice at a density of one particle per unit cell (see Fig.~\ref{fig1}), by using a new numerically exact hybrid quantum Monte Carlo technique which permits us to simulate the ground state of interacting Bose gases in arbitrarily weak or strong periodic potentials. Our main goal is to uncover the physics of the Mott transition in this continuous-space system, going beyond  approximate discrete lattice models. We find a superfluid-to-insulator quantum phase transition, calculate the critical optical lattice intensity as a function of the  $s$-wave scattering length, and compare our results to an experiment performed with an ultracold gas of Cs atoms~\cite{mark}, finding good agreement (see Fig.~\ref{fig1}).

Bosons in an optical lattice are described by the continuous-space Hamiltonian
\begin{equation}
 \label{H}
 \hat{H} = 
 -\frac{\hbar^2}{2m} 
  \sum_{i=1}^N \left[
 \mathbf{\nabla}^2_i + 
 V\left( \mathbf{r}_i\right)
 \right]
 +
 \sum_{i < j} v \left( \mathbf{r}_{ij} \right),
\end{equation}
where $m$ is the atomic mass, the indices $i$ and $j$ label the $N$ particles at positions  $\mathbf{r}_i$, and $\mathbf{r}_{ij} = \mathbf{r}_{i} - \mathbf{r}_{j}$.
The optical lattice potential is $V\left( \mathbf{r}\right) = V_0 \sum_{\alpha = x,y,z}\sin^2\left( \pi \alpha / d \right)$, where $V_0$ is the intensity of the laser beams that create the lattice and $d$ is half  the wave-length.

The third term in Eq.~(\ref{H}) describes the two-body interactions. Different models for the interparticle interactions can be considered, including potentials with a hard~core or with long-range interactions. One specific potential we use is the hard-sphere (HS) potential $v^{\mathrm{HS}}\left(r\right) = \infty$, if $r < a_\mathrm{s}$ and $0$ otherwise. The diameter of the sphere is equal to the $s$-wave scattering length $a_\mathrm{s}$, which is the parameter that characterizes the strength of interactions in dilute gases at low temperature. To quantify the role of other details of the interparticle potential, which become relevant at high scattering energies and at short interparticle distances, we also employ different models with the same $s$-wave scattering length, the negative-power potentials (NP) $v^{\mathrm{NP}} (r)= c /r^{p}$, with $c > 0$ and the integer $p > 3$~\cite{asNP}. 
We do not consider attractive potentials supporting bound states, such as those used to model Feshbach resonances~\cite{chin2010feshbach}, since in this case the possible metastable gaslike phase is not the ground state.

To map out the phase diagram, we employ a new algorithm combining the stochastic integration of variational trial wavefunctions defined on a discrete lattice with an imaginary-time propagation in continuous space. 
The ground-state properties of a homogeneous quantum fluid [$V(\mathbf{r}) = 0$] can be obtained by using the ground-state path-integral Monte Carlo method~\cite{sarsa2000path}, where the expectation value of a generic operator $\hat{O}$ is obtained as
\begin{equation}
\label{expvalue}
 \langle \hat{O} \rangle
= \frac{
\int \mathrm{d}\mathbf{R}\mathrm{d}\mathbf{R}' \Psi^*_\mathrm{T} \left(\mathbf{R}\right)
e^{ - \frac{\beta}{2} \hat{H} }
\hat{O}
e^{ - \frac{\beta}{2} \hat{H} }
\Psi_\mathrm{T} \left(\mathbf{R}'\right)
}{
\int \mathrm{d}\mathbf{R}\mathrm{d}\mathbf{R}' \Psi^*_\mathrm{T} \left(\mathbf{R}\right)
e^{ - \beta \hat{H}  }
\Psi_\mathrm{T} \left(\mathbf{R'}\right)
},
\end{equation}
where $\mathbf{R} = \left(\mathbf{r}_1,\dots,\mathbf{r}_N\right)$. For a sufficiently long imaginary time $\beta$, this expression gives the exact ground-state result provided that the trial wave function $\Psi_\mathrm{T}\left(\mathbf{R}\right)$ has a finite overlap with the ground state. The imaginary-time evolution operator $\exp\left(- \beta/2 \hat{H} \right)$ can be represented in the form of a discretized path integral using the Trotter break up, and the resulting multivariates integral can be evaluated by using standard Monte Carlo sampling techniques~\cite{ceperley1995path,boninsegni2006worm}.
However, the simulation of the strongly correlated regime in a periodic potential with the ground-state path-integral Monte Carlo method is unfeasible with standard trial wave functions. We thus combine the ground-state path-integral Monte Carlo algorithm with the stochastic sampling of variational wave~functions defined on a discrete lattice. The lattice wave function is characterized by the  occupation numbers of a set of orbitals  localized at the nodes of the lattice. Determining the expectation value (\ref{expvalue}) with such trial wave functions requires the sampling of those quantum numbers  together with the continuous-space coordinates $\mathbf{R}$, as explained below.

As a lattice wave function we use a bosonic Gutzwiller ansatz with contact interaction~\cite{rokhsar1991gutzwiller,krauth1992gutzwiller} $\phi\left(\mathbf{l}_1,\dots,\mathbf{l}_N\right) = \exp\left[ -\gamma \sum_{i<j} \delta_{\mathbf{l}_i\mathbf{l}_j} \right]$, where $\delta_{\mathbf{l}_i\mathbf{l}_j}$ is the Kronecker delta, $\gamma$ is a variational parameter, and $\mathbf{l}$ refers to one of the sites of a simple cubic lattice.
For the localized orbitals, we consider a (positive definite) Gaussian approximation of the Wannier functions $w_{\mathbf{l}}(\mathbf{r}) = M^{-1/2}\sum_\mathbf{k} \exp\left(-i\mathbf{k}\cdot\mathbf{l}\right) b_\mathbf{k}(\mathbf{r})$, where $M$ is the number of lattice sites and  $b_\mathbf{k}\left(\mathbf{r}\right)$ is the Bloch state of the lowest band with quasimomentum $\mathbf{k}$.
For a given lattice configuration $\mathbf{L} = \left(\mathbf{l}_1,\dots,\mathbf{l}_N\right)$, the coordinate-space wave function corresponding to $\phi\left(\mathbf{L}\right)$ is $\psi_\mathbf{L}\left(\mathbf{r}_1,\dots,\mathbf{r}_N\right) = \prod_{i=1}^N w_{\mathbf{l}_i}\left(\mathbf{r}_i\right)$, so that the combined trial wave function is $\Psi_\mathrm{T}\left(\mathbf{R}\right) = \sum_\mathbf{L} \phi\left(\mathbf{L}\right) \psi_\mathbf{L}\left(\mathbf{R}\right)$, where the sum goes over all lattice configurations $\mathbf{L}$.
To improve $\Psi_\mathrm{T}$ beyond the contact-interaction ansatz we introduce a Jastrow term $\chi\left(\mathbf{R}\right) = \prod_{i<j} f\left(\mathbf{r}_{ij}\right)$, giving a final trial wave function $\Psi_\mathrm{T}\left(\mathbf{R}\right) = \chi\left(\mathbf{R}\right) \sum_\mathbf{L}\phi\left(\mathbf{L}\right) \psi_\mathbf{L}\left(\mathbf{R}\right)$.
The Jastrow correlation function \cite{jastrow} $f\left(\mathbf{r}\right)$ is set equal to the free-space solution of the two-body problem up to a cutoff distance $r_\mathrm{c}$ and equal to $1$ for $|\mathbf{r}| > r_\mathrm{c}$, as in Ref.~\onlinecite{giorgini1999ground}. With this choice, the short-range correlations due to the hard~core of the interatomic interaction $v\left(\mathbf{r}\right)$ are treated exactly in the trial state $\Psi_\mathrm{T}$, while the long-range correlations can be efficiently reproduced by the imaginary-time propagation. The time evolution also includes the effect of the higher Bloch bands.

To evaluate the right-hand side of Eq.~(\ref{expvalue}), we need to perform a stochastic sampling over the lattice configurations $\sum_\mathbf{L}$ and the particle coordinates $\int \mathrm{d}\mathbf{R}$. This requires a random walk in the combined configuration space $\{\mathbf{L},\mathbf{R}\}$, which we perform by using local Monte Carlo updates on the lattice index $\mathbf{l}_i$ and the space coordinate $\mathbf{r}_i$ of a given particle $i$.

In the present problem, the bosonic statistics plays a central role. The trial state $\Psi_\mathrm{T}$ is by construction symmetric under particle exchange, and the imaginary-time propagation is  symmetrized by performing particle permutations using the worm algorithm~\cite{boninsegni2006worm}. These permutations can efficiently introduce more exotic short-range correlations which are not explicitly included in the trial wave function $\Psi_\mathrm{T}$~\cite{kaplan1982close,capello2007superfluid}.

For some specific  values of $\gamma$, the trial state $\Psi_\mathrm{T}$ reproduces wave functions that have previously been used. When $\gamma \rightarrow 0$, $\Psi_\mathrm{T}$ converges to the ``periodic+Jastrow'' wave-function $\Psi_\mathrm{T}\left(\mathbf{R}\right) \simeq \chi\left(\mathbf{R}\right) \prod_i b_\mathbf{k = 0}\left(\mathbf{r}_i\right)$~\cite{ceperley1979monte} (ignoring an irrelevant normalization coefficient). This wave function accurately describes the ground state of a superfluid Bose gas in a periodic potential. 
In the opposite limit $\gamma \rightarrow \infty$, one obtains a symmetrized state with one particle per site~\cite{gammainfty} (we consider the case $N = M$) $\Psi_\mathrm{T}\left(\mathbf{R}\right) \simeq \chi\left(\mathbf{R}\right) \sum_{\mathbf{P(L)}} \prod_i w_{\mathbf{p}(\mathbf{l}_i)}\left( \mathbf{r}_i \right)$, where the sum runs over the $N!$ permutations $\mathbf{P(L)} = \left(\mathbf{p}(\mathbf{l}_1),\dots,\mathbf{p}(\mathbf{l}_N)\right)$ of $N$ particles in $M=N$ lattice sites. 
In numerical studies of bulk helium-4, a state analogous to this was introduced in order to describe a potential supersolid ground state~\cite{ceperley1979monte,cazorla2009bose}, where superfluidity coexists with diagonal long-range order (in this context it is referred to as ``permanent+Jastrow'', and the Wannier functions are replaced by effective localized orbitals). However, it has never been implemented in other exact quantum Monte Carlo methods such as the diffusion Monte Carlo simulations because of the difficulty of performing random walks in permutation space. 
In the context of the single-band Bose-Hubbard model, the Gutzwiller ansatz $\phi\left(\mathbf{L}\right)$ predicts, in the limit $\gamma \rightarrow \infty$, a quantum phase transition from a superfluid to an insulating solid~\cite{rokhsar1991gutzwiller,krauth1992gutzwiller}. This transition is also seen by exact lattice quantum Monte Carlo simulations~\cite{capogrosso2007phase}. Here we study this transition in the full Hilbert space of the continuum model and investigate deviations from the single-band lattice model.

\begin{figure}[t]
\begin{center}
\includegraphics[width=\columnwidth]{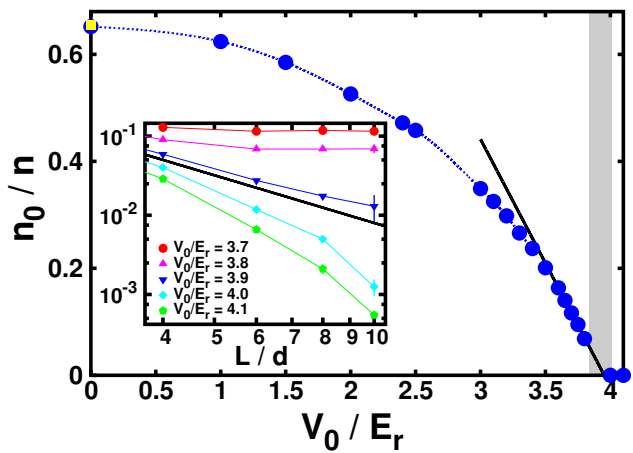}
\caption{(color online). Main panel: Condensate fraction vs optical lattice intensity for HS with $a_\mathrm{s} = 0.3d$ (blue bullets). The black solid line is a linear fit used to extract the critical point $V_0^\mathrm{c}$. The yellow square at $V_0 = 0$ is the diffusion Monte Carlo result of Ref.~\cite{giorgini1999ground}. Inset: Finite-size scaling analysis of $n_0$; $L$ is the simulation-box size. The black solid line is the scaling behavior expected at the critical point: $n_0 \propto L^{-2}$.}
\label{fig2}
\end{center}
\end{figure}

We identify the superfluid phase by measuring the condensate fraction~\cite{notecondensate}, which is efficiently evaluated in the worm algorithm as 
$n_0/n =  N^{-1}\int \mathrm{d}\mathbf{r}
          \lim_{\left|\mathbf{r}-\mathbf{r}'\right|\rightarrow\infty}
\langle   \hat{\Psi}^\dagger\left(\mathbf{r}'\right)
          \hat{\Psi}\left(\mathbf{r}\right)           \rangle$,
where $\hat{\Psi}^\dagger$ ($\hat{\Psi}$) is the creation (annihilation) field operator.
The results for the HS potential with $a_\mathrm{s} = 0.3d$ are shown in Fig.~\ref{fig2}. In the homogeneous limit $V_0 = 0$, we find agreement with the diffusion Monte Carlo prediction of Ref.~\onlinecite{giorgini1999ground} (yellow square). By increasing $V_0$, the condensate fraction decreases monotonically. 
Close to a critical optical lattice intensity (call it $V_0^\mathrm{c}$), $n_0$ approaches zero linearly, consistently with the scaling law of the order parameter in the Bose-Hubbard model $\langle\hat{\Psi}^\dagger(\mathbf{r})\rangle \propto \sqrt{n_0/n}   \propto \left[ (V_0 - V_0^\mathrm{C})/V_0^\mathrm{C} \right]^\beta$, 
where $\beta = 1/2$ (plus logarithmic correction)~\cite{fisher1989boson}. 
With a linear extrapolation to $n_0 = 0$ (black solid line in Fig.~\ref{fig2}), we obtain the critical value $V_0^\mathrm{c}/E_\mathrm{r} = 3.95(10) $, where $E_\mathrm{r} = \hbar^2\pi^2/2md^2$ is the recoil energy. Different extrapolation procedures yield compatible results. 
In order to better identify the quantum phase transition, we perform a finite-size scaling analysis. In the inset in Fig.~\ref{fig2}, we show $n_0/n$ as a function of the simulation-box size $L$ (we employ periodic boundary conditions) for different values of $V_0$. While slightly away from $V_0^\mathrm{c}$ the condensate fraction rapidly saturates to the thermodynamic limit, close to the critical point the scaling of $n_0/n$ approaches the power~law $n_0 \propto L^{-2}$, which is indeed the scaling expected for the order parameter squared~\cite{fisher1989boson}.

\begin{figure}[t]
\begin{center}
\includegraphics[width=\columnwidth]{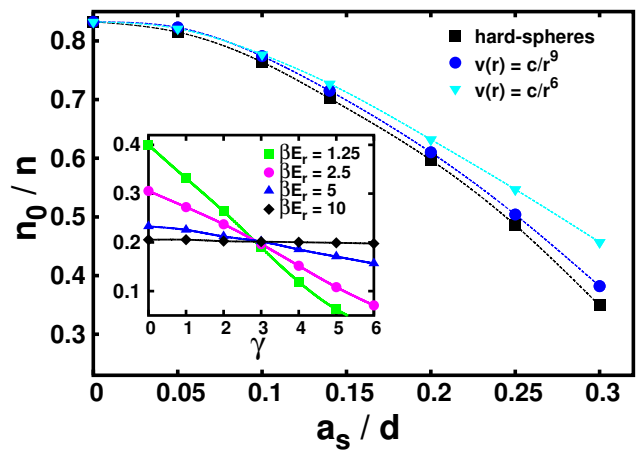}
\caption{(color online). Main panel: Condensate fraction vs s-wave scattering length $a_\mathrm{s}$ for different model potentials. The optical lattice intensity is $V_0 = 3 E\mathrm{r}$. Inset: Condensate fraction as a function of the variational parameter $\gamma$. Here $V_0 = 3.5 E\mathrm{r}$ and $a_\mathrm{s} = 0.3d$. The curves corresponding to different imaginary times $\beta$ cross at the optimal $\gamma$.}
\label{fig3}
\end{center}
\end{figure}

In the vicinity of the critical point $V_0^\mathrm{c}$, the imaginary time $\beta$ required to converge to the ground state scales with the size of the system. Hence, an accurate optimization of $\Psi_\mathrm{T}$ is important to avoid a possible bias due to the variational ansatz. Rather than by minimizing the ground-state energy, this optimization is achieved more conveniently with a finite imaginary-time scaling analysis, as shown in the inset in Fig.~\ref{fig3}. The values of $n_0/n$ obtained with a small imaginary time $\beta$ show a clear dependence on the variational parameter $\gamma$. This dependence vanishes for larger $\beta$. The crossing point of the curves of $n_0/n$ vs $\gamma$ corresponding to different values of $\beta$ provides the optimal value of $\gamma$ which removes the residual finite imaginary-time effects.\\

In the phase diagram of Fig.~\ref{fig1}, we show $V_0^\mathrm{c}$ as a function of $a_\mathrm{s}$. In the region $V_0 < V_0^\mathrm{c}$, the condensate fraction is finite and the system is superfluid. 
The region where $n_0 = 0$ is instead identified as an insulating phase.
The phase boundary obtained with the HS potential (red circles) is compared to that of the Bose-Hubbard model~\cite{capogrosso2007phase} (lower solid gray curve), based on the mapping to single-band models described in Ref.~\cite{jaksch1998cold}. 
For $a_\mathrm{s} \lesssim 0.3d$ and $V_0 \gtrsim 4 E_\mathrm{r}$, the two models follow the same trend, with small discrepancies due to the effect of higher Bloch bands and to the different models for the interatomic potential (HS vs nonregularized delta).
The experimental data~\cite{mark} (black diamonds) are consistent with both the continuum and lattice model but are in better agreement with our continuum  model.
%
%Non-universal effects can be relevant also at small $a_\mathrm{s}$ -but large $V_0$- due to the increased %energy of the single-particle orbitals which enhances the inter-atomic scattering momentum.
%
The yellow dashed curve in Fig.~\ref{fig1} corresponds to a Bose-Hubbard model with occupation-number-dependent parameters due to multiorbital effects, solved within the mean-field approximation (from Ref.~\cite{luehmann}), and clearly simulations going beyond mean-field will be important to compare this model to our continuum results.

In the regime $a_\mathrm{s} \gtrsim 0.3d$, the HS result bends down, as one  approaches the freezing density of hard spheres in free space $V(\mathbf{r}) = 0$~\cite{kalos1974helium} (red vertical arrow). This region of the phase diagram could host more exotic phases, like a quantum-glassy phase eventually coexisting with superfluidity, due to the competition between the spatial  symmetry of the optical lattice (simple cubic) and that of the hard-sphere crystal in free space (face-centered cubic). We leave the investigation of these issues  to future work.\\
In this part of the phase diagram, universality in terms of $a_\mathrm{s}$ is lost and also other details of the interatomic potential become important. We quantify these nonuniversal effects in Fig.~\ref{fig3} (main panel) by comparing the results for $n_0/n$ vs $a_\mathrm{s}/d$ (at fixed $V_0 = 3E_\mathrm{r}$) obtained with different model potentials.
The data corresponding to HS (black squares) and to NP with $n = 9$ (blue circles) agree within $10\%$ up to $a_\mathrm{s} = 0.3d$, indicating a large degree of universality. As expected, the data for NP with $n = 6$ (cyan triangle) show larger deviations. Notice that for $n < 6$ the effective range diverges~\cite{mott1965theory}. 
In the limit $a_\mathrm{s}\rightarrow 0$, $n_0/n$ converges to $n_0/n = 0.833$, which is the square of the overlap between the condensate wave function (the Bloch state with $\bf{k}=0$) and the plane wave with zero momentum.

Our results for the Mott transition in a continuous-space system of bosons with short-ranged interactions can be useful to extend experiments performed with ultracold gases to the regime of strong interactions and weak optical lattices, that is beyond the physics of the Hubbard model. 
While the existence of a quantum phase transition in the single-band model was rigorously proven~\cite{fernandez2006mott}, our findings indicate that quantum critical behavior persists when all Bloch bands are included, in  contrast to the arguments of Anderson~\cite{anderson2009gross,anderson2011ground}.

The hybrid lattice-continuum quantum Monte Carlo method presented in this work can also be extended to fermionic models and be combined  with  fixed-node and  transient-estimate techniques. This will provide a new approach for the simulation of strongly correlated fermionic systems in weak optical lattices.

We acknowledge J. Fr\"olich and H.-P. B\"uchler for fruitful discussions and M. Mark and D.-S. L\"uhmann for providing their data. This work was supported by the Swiss National Science Foundation and a grant by the Army Research Office within the DARPA OLE program.


\begin{thebibliography}{99}
\bibitem{asNP} The scattering length of the NP potential is
$ a_\mathrm{s}=\left( \frac{2m c/\hbar^2}{(p-2)^2} \right)^{1/(p-2)}
\frac{\Gamma[(p-3)/(p-2)]}{\Gamma[(p-1)/(p-2)]}$, where $\Gamma(x)$ is the Gamma function.
\bibitem{jastrow} We impose continuity of $f\left(r\right)$ and of its derivative in $r_\mathrm{c}$.
\bibitem{notecondensate} Strictly speaking, $n_0/n$ is the occupation of the zero momentum state. One has $n_0 \neq 0 \Leftrightarrow n_c \neq 0$, where $n_c$ is the occupation of the Bose-Einstein condensate.
\bibitem{gammainfty} The case $\gamma = \infty$ is simulated with a fictitious finite $\gamma$, measuring observables only when $\mathbf{l}_i\neq \mathbf{l}_j$ for $j\neq i$.
\end{thebibliography}
\end{document}